\documentclass[12pt]{article}
\pdfoutput=1
\usepackage{epsfig}
\usepackage{amsfonts}
\usepackage{amssymb}
\usepackage{xcolor}
\usepackage{amsmath}
\usepackage{graphicx}
\usepackage{cite}
\begin{document}
\begin{center}

{\bf \Large   Dual Conformal Symmetry and Iterative Integrals\vspace{0.3cm} in Six Dimensions} \vspace{1.0cm}

{\bf \large L.V. Bork$^{1,2}$, R.M. Iakhibbaev$^{1}$, D.I. Kazakov$^{1,3}$, \\[0.3cm] and D.M. Tolkachev$^{1,4}$} \vspace{0.5cm}

{\it $^1$Bogoliubov Laboratory of Theoretical Physics, Joint
Institute for Nuclear Research, Dubna, Russia.\\
$^2$Alikhanov Institute for Theoretical and Experimental Physics, Moscow, Russia\\
$^3$Moscow Institute of Physics and Technology, Dolgoprudny, Russia\\ and \\
$^4$Stepanov institute of Physics, Minsk, Belarus}
\vspace{0.5cm}

\abstract{In this article, we continue the investigation of \cite{Lip} regarding iterative properties of dual conformal integrals in higher dimensions. In $d=4$, iterative properties of four and five point dual conformal integrals manifest themselves in the famous BDS ansatz conjecture. In \cite{Lip} it was also conjectured that a similar structure of integrals may reappear in $d=6$. We show that one can systematically, order by order in the number of loops, construct combinations of $d=6$ integrals with $1/(p^2)^2$ propagators with an iterative structure similar to the $d=4$ case. Such combinations as a whole also respect dual conformal invariance but individual integrals may not.}
\end{center}


\section*{Introduction}
Study of the S-matrix (the scattering amplitudes) in supersymmetric gauge theories has revealed
the existence of hidden symmetries and unexpected properties of this type of objects. It appears that the S-matrix often exhibits symmetries that are hidden from the point of view of the standard Lagrangian formulation of the theory. Applying modern methods for computing the on-shell scattering amplitudes, it is possible to learn more about the S-matrix of many theories even without direct reference to their Lagrangians using only symmetry considerations (see, for example, \cite{Henrietta_Amplitudes,Talesof1001Gluons} for review).

Dual conformal invariance of N=4 SYM is a canonical example of such hidden symmetries \cite{Drummond:2008vq}. It
imposes very powerful constraints on the S-matrix of N=4 SYM in the planar limit \cite{Drummond:2008vq,Drummond_Yangian}. One of the manifestations of these powerful constraints is a very limited set of master integrals contributing to the n-point amplitudes at the low loop level \cite{BDS1,Generalized_unitarity}. In addition, it was pointed out that such integrals have an  iterative structure \cite{BDS1}. These observations culminated in the famous BDS conjecture \cite{BDS1} for the MHV planar amplitudes in N=4 SYM,
which indeed gives a correct all loop expression  for the four and five point amplitudes \cite{ampWLduality5}.

In \cite{Lip}, it was conjectured that there may exist some $d=6$ gauge theories with the S-matrix elements (amplitudes) in the planar limit given by $d=6$ dual (pseudo)conformal integrals, similar to the N=4 SYM case. The distinct feature of such integrals, compared to the $d=4$ case, is the presence of $1/(p^2)^n$ propagators ("dots on the lines").
It was also conjectured in \cite{Lip} that in the strong coupling regime the amplitudes in
a theory like this can also be evaluated similar by the $d=4$ N=4 SYM case \cite{ampWLduality1}
i.e. they should be identified with minimal surfaces in
$\text{AdS}_7$. Under certain assumptions the behaviour of four point amplitude in such a $d=6$ theory should be nearly identical to the $d=4$ BDS ansatz and  for the normalized colour ordered amplitude $M_4=A_4/A_4^{tree}$ can be written as
\begin{eqnarray}\label{BDS_Intro}
 M_4\sim exp\left(\mathcal{S}_{s}+\mathcal{S}_{t}
+\frac{\gamma_{cusp}(g_6)}{8} L^2\left(\frac{s}{t}\right)\right),
\end{eqnarray}
where $s,t$ are the  standard Mandelstam variables, $g_6$ is the  properly normalized coupling constant (see \cite{Lip} for discussion) in six dimensions,  $L(x)\equiv \log(x)$ and  $\mathcal{S}_{Q^2}$ are given by
\begin{equation}
\mathcal{S}_{Q^2}=\sum_{l=1}\frac{g^l_6}{4}\left(\frac{\mu^2}{-Q^2}\right)^{l\epsilon}\left(\frac{\gamma_{l}}{(l\epsilon)^2}+\frac{2G_{l}}{l\epsilon}\right),
\end{equation}
where the coefficients $\gamma_{l}$ and $G_{l}$  define the so-called cusp $\gamma_{cusp}(g)=\sum_lg^l\gamma_{l}$ and collinear $G(g)=\sum_lg^lG_{l}$ anomalous dimensions \cite{BDS1}. The results of \cite{Lip} claim that $\gamma_{cusp}(g)$ may be identical  for the $d=4$ and $d=6$ cases, while $G(g)$ is different.

This suggests that  some
iterative pattern should exist for appropriately chosen
combinations of four point $d=6$ dual (pseudo)conformal integrals
similar to their $d=4$ counterparts. One can also speculate that this six dimensional theory may be in fact mysterious $(2, 0)$ SYM \cite{Lip}. It is worth mentioning that $d=6$ gauge theories, among  $(2, 0)$ SYM one, where extensively studied in recent years for different reasons and from different perspectives. In this regard we want to mention \cite{20theor1,20theor2,20theor3,20theor4,20theor5} and \cite{d6Amp1,d6Amp2,d6Amp3} and also \cite{d6UV1,d6UV2,d6UV3,d6UV4,d6UV5,d6UV6,B1,B2,B3}. 

Inspired by \cite{Lip} we investigate possible
iterative relations between the four point 6d dual (pseudo)conformal
integrals listed in \cite{Lip} at the two and three-loop level.
The article is organized as follows: in section 1, we
briefly discuss dual conformal symmetry and show how it manifests itself in two types of regularization.

In section 2, we discuss how one can explicitly evaluate the four point dual (pseudo) conformal
integrals at the one and two-loop level. We also present the explicit results for the four point dual conformal integrals with the double box topology listed in \cite{Lip}. In addition, we also discuss some higher loop results.

In section 3, we investigate possible iterative relations between the one and two-loop
$d=6$ dual conformal integrals listed in \cite{Lip}. In general, we did not find any iterative relations
among these integrals except for the particular subset evaluated at the  kinematic point $s=t=-Q^2$. This observation however allows us to present the procedure based on the $d=4$ expansion of the BDS four point amplitude in terms of the master integrals, which allows one to construct combinations of $d=6$ integrals
with  the iterative structure. These combinations respect dual conformal invariance if an appropriate regularization is chosen, but individual integrals may not be dual conformal invariant.

In conclusion, we sum up our results and in appendix we list the answers for $d=6$ integrals mentioned in the text.

\section{Dual Conformal Symmetry}

Dual conformal symmetry is conformal symmetry in momentum space.
It can be realized as follows. One introduces the
coordinates $x^\mu_i$ that are defined as
\begin{equation}
p_{i}^\mu=x^\mu_i-x^\mu_{i+1},
\end{equation}
and then considers the action  of the standard conformal generators
on these new variables $x^\mu_i$. For example, the action
of inversion $I$ on $x^\mu_i$ and $x^2_{i j}=(x_i-x_j)^2$ is given by:
\begin{equation}
I[x^\mu_i] = \frac{x^\mu_i}{x^2_i},~I[x^2_{i j}] = \frac{x^2_{ij}}{x^2_i x^2_j}.
\end{equation}

\begin{figure}[ht]
 \begin{center}
 %\leavevmode
  \epsfxsize=12cm
 \epsffile{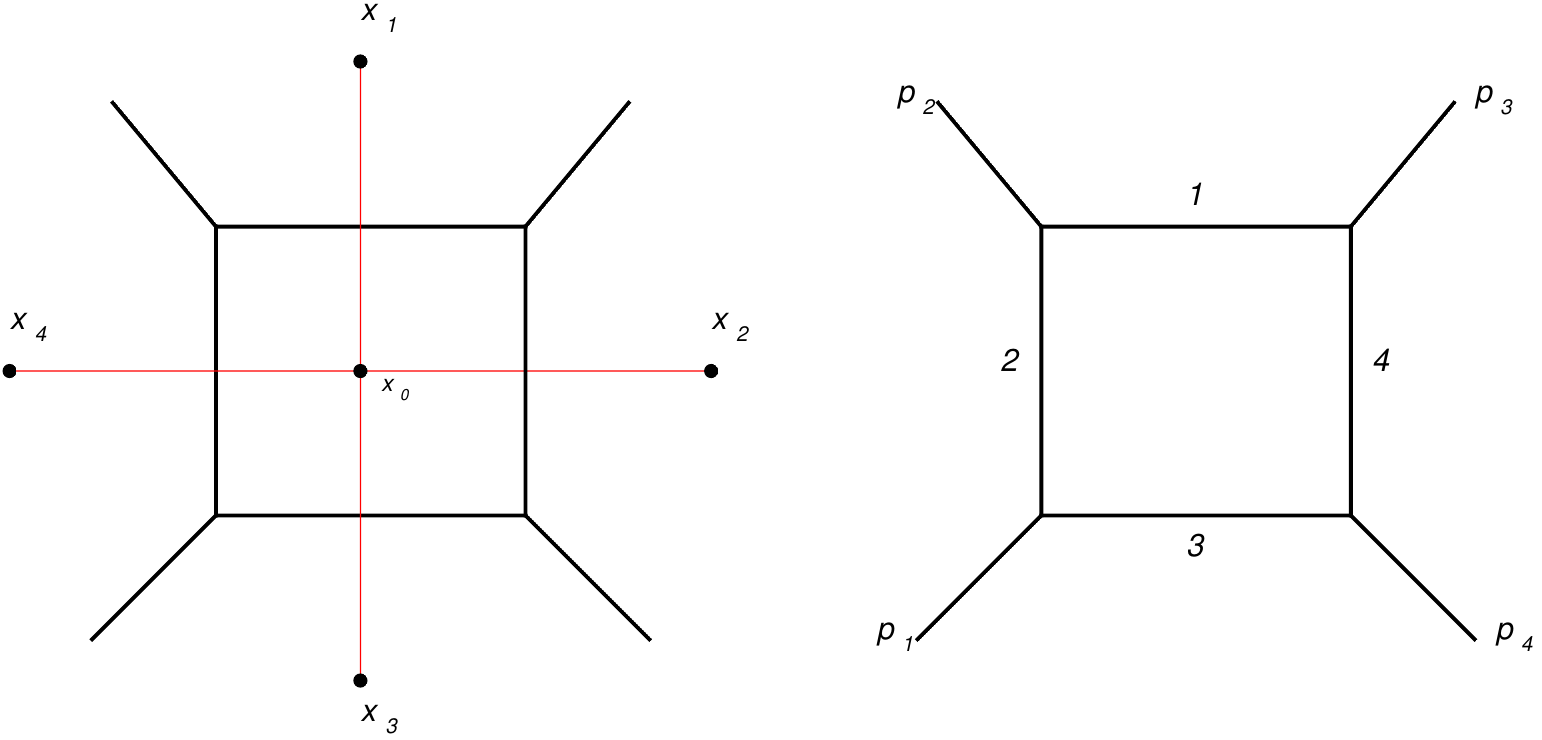}
 \end{center}\vspace{-0.2cm}
 \caption{Scalar box integral in dual and standard variables. Numbers on the propagators numerate corresponding alpha parameters.}\label{d4_Box}
 \end{figure}

To see how this symmetry manifests itself at the level of Feynman integrals, let us consider the 1-loop box diagram in $d$ dimensions which is given by (see fig. \ref{d4_Box})

\begin{equation}
\label{singleboxdualvariables4}
Box_4^{d}(s,t)=\int d^d k\frac{ (p_1+p_2)^2(p_2+p_3)^2}{k^2(k+p_1)^2  (k+p_1+p_2)^2 (k-p_4)^2} =\int d^d x_0 \frac{x^2_{13}x^2_{24}}{x^2_{10} x^2_{20} x^2_{30} x^2_{40}}.
\end{equation}
Having in mind that the measure of integration  transforms under $I$ as
\begin{equation}
d^d x_i\rightarrow \frac{d^d x_i}{(x^2_i)^d},
\end{equation}
it is easy to see that this integral is invariant with respect to $I$ and other
conformal transformations of $x_{0i}$ and $x_{ij}$ when $d=4$.
When $x_{ii+1}^2\neq0$, which is equivalent to $p_i^2\neq0$, this integral is finite and can be evaluated exactly \cite{Usyukina1,Usyukina2} (see appendix A). Dual conformal invariance guarantees that  the function $Box_4^{d=4}(s,t)=f(u,v)$ depends on two conformal cross ratios
\begin{equation}\label{Cross1}
u=\frac{x^2_{12}x^2_{34}}{x^2_{13}x^2_{24}},~
v=\frac{x^2_{14}x^2_{23}}{x^2_{13}x^2_{24}}.
\end{equation}
For example, one can see that the action of the dual conformal boost $K^{\nu}$ defined by
\begin{eqnarray}\label{DualConfBoost}
&&K^{\nu}=\sum_{i=1}^4\left(2x^{\nu}_i(x_i\partial_i)-x^2_i\partial^{\nu}_i\right).
\end{eqnarray}
on $Box_4^{d=4}(s,t)$ is zero:
\begin{eqnarray}\label{DualConfBoostAction}
&&K^{\nu}~Box_4^{d=4}(s,t)=0,
\end{eqnarray}
Obviously, higher loop integrals can also be dual conformal invariant. See fig.\ref{d4_DBox}  for the two loop $d=4$ example.

If $x_{ii+1}^2\equiv p_{i}^2=0$ (i.e. external momenta are on-shell), the $d=4$ one loop box integral as well as its higher loop counterparts are IR divergent. These
divergences can be regularized, for example, by dimensional regularization in
$d=4-2\epsilon$ which, however, will spoil dual conformal invariance. The divergent and finite parts of the box function in this case can be found in Appendix.  The answer in all orders in $\epsilon$ is  known in terms of hypergeometric functions. Such regularized integrals with the on-shell external momenta are usually
called \emph{pseudo dual conformal integrals}. In this on-shell regime the notion of dual conformal invariance is somewhat obscure; however, for small $\epsilon$ the breaking of dual conformal invariance should be under control in some way. Indeed, in the case of N=4 SYM, one can show \cite{ampWLduality5} that at least for appropriate combinations of (pseudo) dual conformal integrals, which give the amplitude $M_4^{d=4-2\epsilon}$, while
\begin{equation}
K^{\nu}~M_4^{d=4-2\epsilon}\neq 0,
\end{equation}
one can consider $log\left[M_4^{d=4}\right])$ instead of $M_4^{d=4}$ and split it
into divergent $\mathcal{S}_s+\mathcal{S}_t$ and finite parts $F_4\sim L^2(s/t)$. Then
for the finite part $F_4$ one has
\begin{eqnarray}\label{DualConformalAnomaly1}
K^{\nu}~F_4=\gamma_{cusp}(g)\sum_{i=1}^4x^{\nu}_iL\left(\frac{x^2_{i,i+2}}{x^2_{i-1,i+1}}\right).
\end{eqnarray}
Other dual conformal generators acts regularly on $M_4$.
This relation can be interpreted as (anomalous) Ward identities and is an analogue of (\ref{DualConfBoostAction}) in
the case when the dual conformal invariance is broken by the IR regulator.

Concluding this section, we consider another way to regularize four point
dual conformal integrals, which we will find useful later. Let us return to the one loop $d=4$ box example.
One can \cite{Henn1} introduce two sets of coordinates: the $d$ dimensional $x_i$, defined as in the previous case as
\begin{eqnarray}
	p_{i}^{\nu}=(x_i-x_{i+1})^{\nu},
\end{eqnarray}
and  the $d+1$ dimensional ones $\hat{x}_i$ defined as
\begin{eqnarray}
	\hat{x}_i^{\nu}=x_i^{\nu},~\hat{x}_i^{d+1}=m_i,
\end{eqnarray}
which transform under dual conformal inversions as
\begin{eqnarray}
	\hat{I}[\hat{x}_{i}^{\nu}]=\frac{\hat{x}_{i}^{\nu}}{x_i^2},
	~\hat{I}[m_i]=\frac{m_i}{x_i^2},~\hat{I}[\hat{x}_{ij}^2]=\frac{\hat{x}_{ij}^2}{x_i^2x_j^2}.
\end{eqnarray}
The square of $\hat{x}_{ij}$ is then given by $\hat{x}_{ij}^2=x_{ij}^2+(m_i-m_j)^2$.
So in this way we can consider massive propagators $1/(p^2+m^2)$ as massless once but in a higher dimensional theory \cite{Henn1}.
We then can rewrite the $d=4$ box integral as
\begin{eqnarray}
Box_4^{d=4}(s,t,m_i)=\int d^5 x_0~ \delta(x_0^5)\frac{\hat{x_{13}^2}\hat{x_{24}^2}}{\hat{x}_{10}^2\hat{x}_{20}^2\hat{x}_{30}^2\hat{x}_{40}^2},
\end{eqnarray}
which in the standard notation is equivalent to
\begin{eqnarray}\label{MassiveBox}
Box_4^{d=4}(s,t,m_i)=\int d^4k \frac{[(p_1+p_2)^2+(m_1-m_3)^2][(p_2+p_3)^2+(m_2-m_4)^2]}{[k^2+m_1^2][(k+p_1)^2+m_2^2][(k+p_1+p_2)^2+m_3^2][(k-p_4)^2+m^2_4]}.\nonumber\\
\end{eqnarray}
This integral is IR finite and is still invariant under most of the (dual)conformal transformations (inversions, rotations) in $d+1=5$ dimensions\footnote{Translation symmetry however is broken down to  $d=4$ dimensions.} \cite{Henn1}.
For example, the generator of the dual conformal boost  in this case is given by
\begin{equation}\label{DualConfBoost_1}
K^{\nu}=\sum_{i=1}^4\left(2x^{\nu}_i(x_i\partial_i)+2x^{\nu}_im_i\partial_{m_i}
-(x^2_i+m_i^2)\partial^{\nu}_i\right),
\end{equation}
so that
\begin{equation}\label{DualConfBoost_1}
K^{\nu}~Box_4^{d=4}(s,t,m_i)=0.
\end{equation}

As a consequence of this invariance, integral (\ref{MassiveBox}) becomes a function
of only two variables \cite{Henn1}
\begin{equation}\label{DualConfRaitMassive}
u=\frac{m_1m_3}{\hat{x}_{13}^2},~
v=\frac{m_2m_4}{\hat{x}_{24}^2}.
\end{equation}
Note that these arguments however are different from the $p_i^2 \neq 0$ case.
The explicit form of this function can also be found in Appendix for the case of small identical $m_i$.

\begin{figure}[t]
 \begin{center}
 %\leavevmode
  \epsfxsize=10cm
 \epsffile{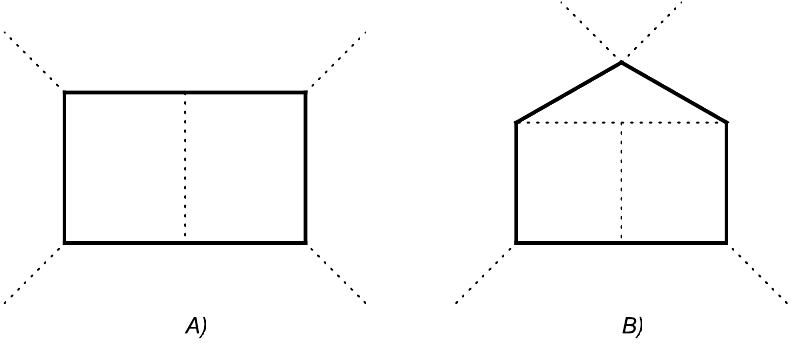}
 \end{center}\vspace{-0.2cm}
 \caption{Example of four point $d=4$ dual conformal integrals in the Higgs regularisation (all regulator masses are equal) \cite{Henn2}. The solid lines correspond to the massive propagators and the dashed lines correspond to the massless ones. Integral A) has an equivalent in dimensional regularisation, integral B) is unique to the Higgs one - it is proportional to $m$. }\label{HiggsReg}
 \end{figure}

The regularization with masses is usually called the Higgs regularization because originally it was introduced in the context of N=4 SYM and the masses in the propagators were related to the vev's of the N=4 SYM scalars \cite{Henn1}.
The higher loop generalizations of this regularization can also be considered at least for $n=4,5$ external legs (see fig. \ref{HiggsReg}).

It is interesting to compare the action of the dual conformal boost generator on the amplitude $M_4^{d=4}$ in such regularization with relation (\ref{DualConformalAnomaly1}), since formally we now have
\begin{equation}
K^{\nu}~M_4^{d=4}\Big{|}_{Higgs~reg.}= 0.
\end{equation}
In the small mass limit, $Log(M_4^{d=4})$ in the Higgs regularization can also be split into divergent and finite parts, and the action of $K^{\nu}$ on the divergent part will produce the term identical to the rhs of (\ref{DualConformalAnomaly1}), so that relation (\ref{DualConformalAnomaly1}) will hold in this regularization as well~\cite{Henn1}.

\section{Four point (pseudo)dual conformal integrals in various dimensions}
As was discussed in the introduction the combinations of (pseudo)dual conformal integrals
in $d=4-2\epsilon$ dimensions possess some remarkable iterative properties \cite{BDS1,BDS4loop,BDS5loop,BDS67loop}. Namely, the four and five point amplitudes
in N=4 SYM can be expressed via such integrals in the weak coupling regime, and the divergent and finite parts of higher loop corrections exponentiate. This means that effectively the amplitudes can be expressed via the one loop contribution and some functions of the coupling constant. These relations are manifested in the famous BDS ansatz  \cite{BDS1}. The results of \cite{Lip} suggest that similar relations may hold for $d=6-2\epsilon$ (pseudo)dual conformal invariant integrals with $n=4$ external legs. We will investigate this conjecture in more detail in the next section while here we concentrate on evaluation of such integrals.

\subsection{Dual conformal bubbles, triangles and boxes at one loop in various dimensions}

Let us consider the case of $n=4$ external legs \cite{Lip} in $d>4$. To get dual (pseudo)confor\-mal invariant integrals in this case, as it was pointed out in \cite{Lip}, one has to consider integrals with dots on the lines, i.e. introduce the propagators  of type $1/(p^2)^2$, etc.
This, however, effectively rules out the bubble and triangle topologies in $d=6$ since
such integrals give constant contributions, that can be associated with the scheme
dependence in the full combination of such integrals if one is interested in BDS-like exponent structures.

This leaves us with one possible topology at the one-loop level, namely, the box-like integral. At $d=6$ dual (pseudo)conformal symmetry uniquely fixes powers of the propagators and the overall coefficient and gives
\begin{equation}\label{d6Box}
Box_4^{d=6}(s,t) = \int d^6 x_0 \frac{x^4_{13}x^2_{24}}{(x^2_{10})^2 x^2_{20} (x^2_{30})^2 x^2_{40}}.
\end{equation}

If $p_i^2=0$ this integral is IR divergent and some regularization is required. If one adopts dimensional regularization, this integral can be evaluated by means of the Mellin-Barnes representation \cite{Smirnov_book_2}. Introducing the function (let us remind that $x_{13}^2=(p_1+p_2)^2=s$ and $x_{24}^2=(p_2+p_3)^2=t$)
\begin{equation}
\mathcal{I}(s,t,\nu_1,\nu_2,\nu_3, \nu_4,d)= \int\frac{ d^d k}{((k)^2)^{\nu_1} ((k+p_1)^2)^{\nu_2} ((k+p_1+p_2)^2)^{\nu_3}  ((k-p_4)^2)^{\nu_4} },
\end{equation}
which is related to $Box_4^{d=6}(s,t)$ as
\begin{equation}
Box_4^{d=6}(s,t)=st~\mathcal{I}(s,t,1,2,1,2,6-2\epsilon),
\end{equation}
one can obtain for $\mathcal{I}$ the  one fold Mellin-Barnes representation:
\begin{equation}
\mathcal{I}(s,t,\nu_1,\nu_2,\nu_3, \nu_4,d)=i\frac{(-1)^\nu \pi^{d/2} s}{\Gamma(d-\nu)\prod^4_{i=1}t^{\nu-d/2}} \int^{+i \infty}_{-i\infty} \frac{dz}{2 \pi i} I(s/t,z)
\end{equation}
with
\begin{equation}
\begin{split}
  I(s/t,z) = \left(\frac{s}{t}\right)^z \Gamma(-z) \Gamma(\nu-d/2+z)\Gamma(\nu_1+z)\Gamma(\nu_3+z) \\ \Gamma(d/2-\nu_{134}-z) \Gamma(d/2-\nu_{123}-z).
\end{split}
\end{equation}
This integral in its  turn can be evaluated\footnote{Thought the paper we used the Mathematica packages \texttt{AMBRE.m} \cite{AMBRE} to obtain the Mellin-Barnes representation for a given integral and \texttt{MB.m} \cite{Czakon_MB_auto} for evaluating the Mellin-Barnes integral. We also used \texttt{MBresolve.m} \cite{Smirnov_Resolution_of_Singularities} -- the additional code for \texttt{MB.m} which implements another strategy of evaluating the Laurent series and  \texttt{MBSums.m} \cite{MBsums} for the series summation.} in the form of a series in $\epsilon$  and the result for $\nu_1=1,\nu_2=2,\nu_3=1, \nu_4=2,d=6-2\epsilon$ is given by:
\begin{equation}
Box_4^{d=6}(s,t)=\left(\frac{\mu^2}{s}\right)^{\epsilon}\sum_{n=2}^{-\infty}\frac{c_n(s/t)}{\epsilon^n},
\end{equation}
with ($x=-s/t$)
\begin{equation}
c_2(x)=4,~c_1(x)=2L(x)-2,\ldots
\end{equation}
(see Appendix for the result up to $\epsilon^1$ terms). These results were first obtained in \cite{Lip}.

It is interesting to note the connection between this $d=6-2\epsilon$ dual  (pseudo)conformal integral and the  standard $d=4-2\epsilon$ box integral discussed in the previous section.
Indeed, one can see that the $\alpha$-representation for the function $\mathcal{I}(s,t,\nu_1,\nu_2,\nu_3, \nu_4,d)$ has the form (see fig.\ref{d4_Box}):
\begin{equation}
\label{alpha1}
\begin{split}
\mathcal{I}(s,t,\nu_1,\nu_2,\nu_3, \nu_4,d)=(-1)^\nu \frac{\pi^{d/2} e^{i \frac{\pi}{2} (\nu+h(1-d/2))}}{\prod^4_{i=1}\Gamma(\nu_i)} \int^\infty_0 \prod^4_{i=1} d \alpha_i \; \prod^4_{i=1}\alpha^{\nu_i-1}_i \mathcal{U}^{-d/2} e^{i \mathcal{V}/{\mathcal{U}}}
\end{split}
\end{equation}
with $\nu=\sum_i \nu_i$, and
\begin{eqnarray}
\mathcal{V}&=&t\alpha_1\alpha_3+s \alpha_2 \alpha_4\\
\mathcal{U}&=&\alpha_1+\alpha_2+\alpha_3+\alpha_4
\end{eqnarray}
Then one can see that the following relations hold:
\begin{equation}
\label{relbd}
\begin{split}
 - \pi  \frac{\partial}{\partial t}\mathcal{I}(s,t,1,1,1, 1,d)= \mathcal{I}(s,t,2,1,2,1,d+2), \\ \; - \pi  \frac{\partial}{\partial s}\mathcal{I}(t,s,1,1,1, 1,d)= \mathcal{I}(t,s,1,2,1,2,d+2),
\end{split}
\end{equation}
so for the box integrals  we get
\begin{equation}\label{1loopd4d6relation}
Box_4^{d=6}(s,t)=-\pi\left(t \frac{\partial}{\partial t}-1\right)Box_4^{d=4}(s,t)
\end{equation}
Due to the $s,t$ symmetry of the $d=4$ box integral a similar result can also be obtained by
taking the partial derivative with respect to $s$.

For $d=4-2\epsilon$ this relation connects all the terms of $\epsilon$ expansion of the dual (pseudo)con\-for\-mal box integral in $d=4-2\epsilon$ and the  integral  in $d=6-2\epsilon$ dimensions.
%See fig for graphical representation of this relation.
We have also verified that this relation holds by explicit evaluation of several first terms of $\epsilon$ expansion for $d=4$ and $d=6$ MB-integrals. This relation explains similarities and discrepancies observed by \cite{Lip} between the $d=6$ conjectured one loop amplitudes and the $d=4$ N=4 SYM amplitude.

Note that this relation does not hold only for dimensional regularized integrals but for the Higgs regularization as well. Indeed the $\alpha$-parametrization for the box integral with massive propagators has a similar form but with slightly modified exponent
\begin{equation}
\label{alpha_massive}
\begin{split}
\mathcal{I}(s,t,m^2,\nu_1,\ldots, \nu_4,d)=(-1)^\nu \frac{\pi^{\frac{d}{2}} e^{i \frac{\pi}{2} (\nu+h(1-d/2))}}{\prod^4_{i=1}\Gamma(\nu_i)} \int^\infty_0 \prod^4_{i=1} d \alpha_i \; \prod^4_{i=1}\alpha^{\nu_i-1}_i \mathcal{U}^{-d/2} e^{i \mathcal{V}/{\mathcal{U}}- im^2\sum_i\alpha_i}
\end{split}
\end{equation}
From this relation we can see that the derivative with respect to $t$  shifts the dimension and powers of propagators identical to the massles case.

\subsection{Dual conformal two loop double boxes}

\begin{figure}[ht]
 \begin{center}
 %\leavevmode
  \epsfxsize=10cm
 \epsffile{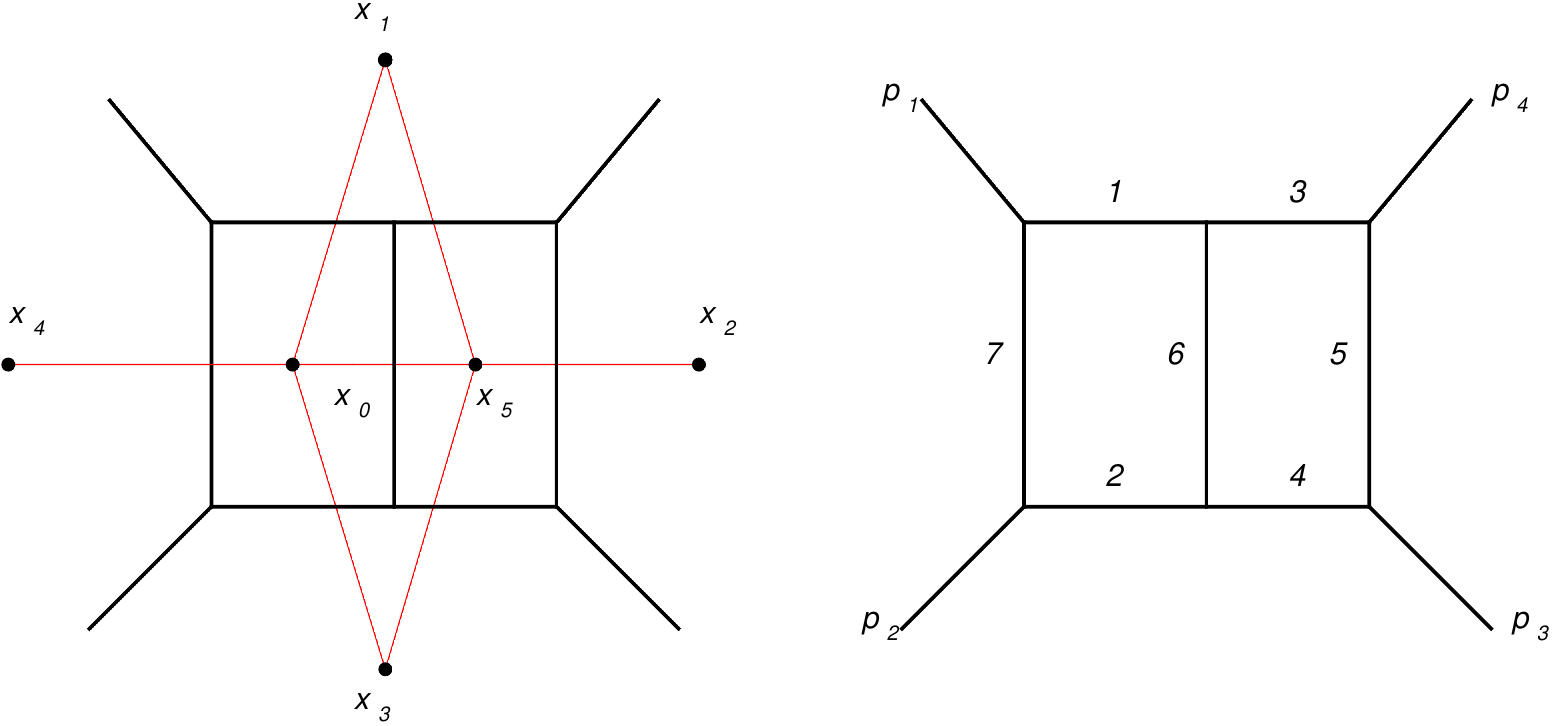}
 \end{center}\vspace{-0.2cm}
 \caption{Double box integral in the dual and standard variables. The numbers on the propagators numerate the corresponding alpha parameters.}\label{d4_DBox}
 \end{figure}

Let us now consider two loop four point dual (pseudo)conformal integrals in $d>4$ (see fig.\ref{d4_DBox}).

In $d=6-2\epsilon$, if one excludes the integrals with bubble and triangle subgraphs \cite{Lip}, the
remaining dual conformal four point integrals are reduced to the form:
\begin{equation}
\label{eq::db61}
DBox_4^{d=6}(s,t)=\int d^D x_0 \; d^D x_5 \frac{(x_{13}^2)^{\alpha} (x_{24}^2)^\beta}{(x_{01}^2)^{\nu_2} (x_{02}^2)^{\nu_1}(x_{04}^2)^{\nu_3} (x_{05}^2)^{\nu_7}(x_{53}^2)^{\nu_5} (x_{54}^2)^{\nu_4} (x_{52}^2)^{\nu_6}},
\end{equation}
where all possible values of $\alpha,\beta$ and $\nu_1,\ldots,\nu_6$  are presented in table (\ref{table1}).
\begin{table}[h!]
\label{table1}
\centering
\caption{The list of indices for independent double box diagrams in six dimensions from \cite{Lip} }
\begin{tabular}{|l|l|l|l|l|l|l|l|l|l|l|l|l|l|l}
\hline
N&$\nu_1$  &  $\nu_2 $ &$ \nu_3 $&$ \nu_4$ &$ \nu_5$&$\nu_6$&$\nu_7$& $\alpha$&$\beta$  \\ \hline  \hline
 (1)& 1 &2  &  1&  1&2 &1  &2  & 2  &2\\ \hline
(2) & 2 &1  & 2 & 2 & 1& 2 & 1 & 1 &4 \\  \hline
(3) & 1 & 1 &1  &1  & 1& 1 & 3 &1  &2 \\ \hline
(4)  & 2 & 1 & 1 & 2 &1 &1  & 2 &  1&3 \\ \hline
(5) & 1 & 1 &2  &1  & 1 &2  & 2 & 1 & 3\\ \hline
(6)& 3 &  1&1  &3 & 1 &1  & 1 & 1 & 4\\  \hline
(7)& 1 & 1 & 3 & 1 &1 & 3 &1  & 1 & 4\\   \hline
(8)& 1 & 2 & 2 &1 & 2 & 2 &1  &2  & 3\\   \hline
(9)& 2 &  2&  1& 2&2 & 1 &1  &2  & 3 \\  \hline
(10)& 1 & 3 & 1 & 1 &3  & 1 &1 & 3 & 2\\  \hline
\hline
\end{tabular}
\end{table}
In the case when $p_i^2=0$, all these integrals are IR divergent. In dimensional regularization
they can be evaluated using  the Mellin-Barnes representation. Defining the general double box integral as (here as usual $x_{13}^2\equiv s$ and $x_{24}^2\equiv t$, and index $N$ corresponds to particular values of $\nu_1,\ldots,\nu_7$ and $\alpha,\beta$ form the table)
\begin{equation}
DBox_4^{d=6,N}(s,t)=(s)^{\alpha} (t)^\beta~\mathcal{I}_2(s,t,\nu_1,\ldots,\nu_7,6-2\epsilon),
\end{equation}
it is possible to get the MB 4-fold representation for $\mathcal{I}_2$:
%s^{-(\sum_i \nu_i-d-\alpha_1)
\begin{equation}
\mathcal{I}_2(s,t,\nu_1,..,\nu_7,d)=\frac{s^{2\epsilon}}{\Gamma(d-\sum_{i=2,4,5,6,7} \nu_i)}\frac{(-1)^{\sum_i \nu_i}\pi^{d/2}}{\prod_{i=2,4,5,6,7} \Gamma(\nu_i)} \\ \int \prod^4_{i=1} \frac{d z_i}{(2 \pi i)^4}I_2(d,s/t,\nu_1,..z_1,..),
\end{equation}
where
\begin{multline}
I_2(t/s,\nu_1,..z_1,..,d)=\left(t/s\right)^{z_1} \Gamma(\nu_2 + z_1) \Gamma(-z_1)\Gamma(z_2 + z_4) \Gamma(z_3 + z_4) \Gamma(-z_2 - z_3 - z_4) \Gamma(
  \nu_7 + z_1 - z_4)  \\  \Gamma(
  \nu_1 + \nu_2 +\nu_3 -d/2 + z_4)  \Gamma(d/2 -\nu_3 - \nu_2 + z_3)    \Gamma(
  \nu_5 + z_1 + z_2 + z_3 + z_4)   \Gamma(
  d/2 - \nu_1 - \nu_2 + z_2) \\ \Gamma(
  d/2 - \nu_4 - \nu_5 - \nu_7 - z_1 - z_3) \Gamma( d/2 - \nu_5 - \nu_6 - \nu_7 - z_1 - z_2) \Gamma(\nu_4 + \nu_5 + \nu_6 + \nu_7 -d/2 + z_1 - z_4).
\end{multline}
We evaluated several first terms of $\epsilon$ expansion of each integral from table (\ref{table1}). Here we present the result for N=1, the other cases can be found in Appendix.
\begin{equation}
DBox_4^{d=6,1}(s,t)=\left(\frac{\mu^2}{s}\right)^{2\epsilon}\sum_{n=2}^{-\infty}\frac{c_n(s/t)}{\epsilon^n},
\end{equation}
with ($x=-s/t$)
\begin{equation}
c_4(x)=-4,~c_3(x)=-5L(x)+5,\ldots.
\end{equation}

\begin{figure}[ht]
 \begin{center}
 %\leavevmode
  \epsfxsize=8cm
 %\epsffile{d4_d6_correspondance_s.eps}
 \epsffile{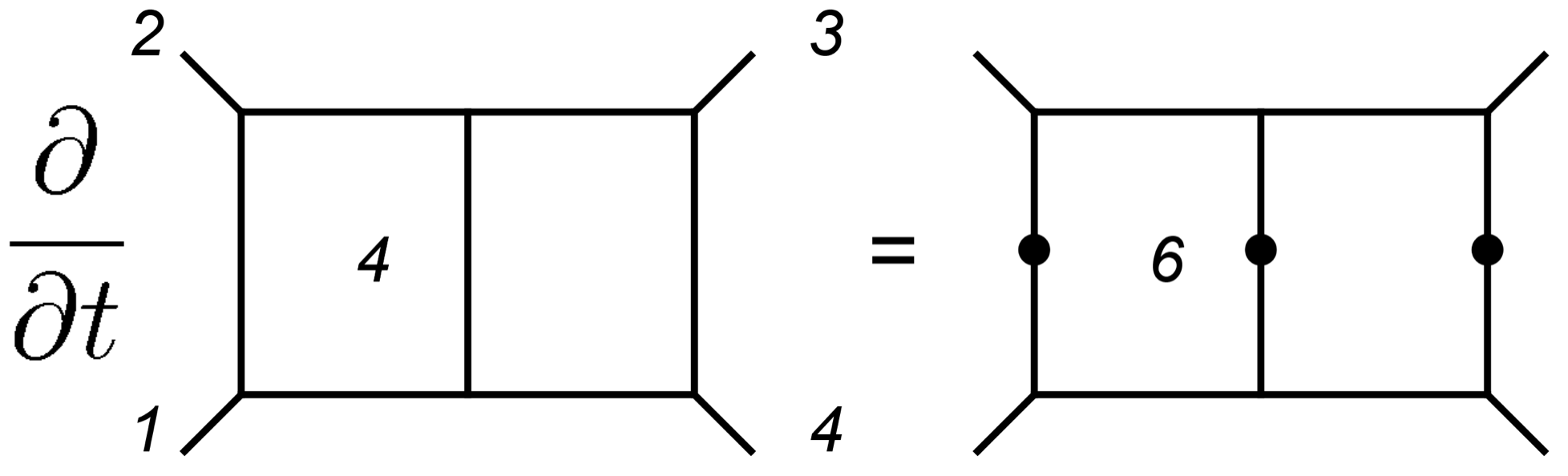}
 \end{center}\vspace{-0.2cm}
 \caption{Relation between two loop four point $d=4$ and $d=6$ double box dual conformal integrals (their $\mathcal{I}$ functions). Numbers in the diagram indicate dimensionality of the corresponding integrals.}\label{ds_DBox}
 \end{figure}

Note that similar to the one loop case, for the $N$=1 case the relation between the
$DBox_4^{d=6,1}(s,t)$ double box and the standard $d=4$ double box $DBox_4^{d=4}(s,t)$ also holds.
Indeed, the $\mathcal{U}$ and $\mathcal{V}$ polynomials in the $\alpha$-representation of $DBox_4^{d=6,1}(s,t)$ now have the form (here $\alpha_2,\alpha_5,\alpha_7$ are $\alpha$ parameters associated with vertical rungs of the horizontal bouble box):
\begin{eqnarray}\label{AlphaPolyDBox}
\mathcal{V}&=&(\alpha_1\alpha_6(\alpha_3+\alpha_4+\alpha_5)+\alpha_3\alpha_4(\alpha_1+\alpha_6+\alpha_7)+\alpha_2(\alpha_1+\alpha_3)(\alpha_6+\alpha_4))s+\alpha_2\alpha_5\alpha_7 t,\nonumber\\
\mathcal{U}&=&(\alpha_1+\alpha_2+\alpha_3)(\alpha_3+\alpha_4+\alpha_5)+\alpha_6(\alpha_1+\alpha_2+\alpha_3+\alpha_4+\alpha_5+\alpha_6+\alpha_7),
\end{eqnarray}
and we can see that
\begin{equation}
\label{relbd}
\begin{split}
 - \pi \frac{\partial}{\partial t}\mathcal{I}(s,t,1,1,1,1,1,1,1,d)= \mathcal{I}(s,t,1,2,1,1,2,1,2,d+2),
\end{split}
\end{equation}
which gives us the  relation identical to the one loop one (see fig. \ref{ds_DBox}) :
\begin{equation}\label{2loopd4d6relation}
DBox_4^{d=6,1}(s,t)=-\pi\left( t  \frac{\partial}{\partial t}-1\right)DBox_4^{d=4}(s,t).
\end{equation}

\begin{figure}[ht]
 \begin{center}
 %\leavevmode
  \epsfxsize=16cm
 %\epsffile{d4_d6_correspondance_t.eps}
 \epsffile{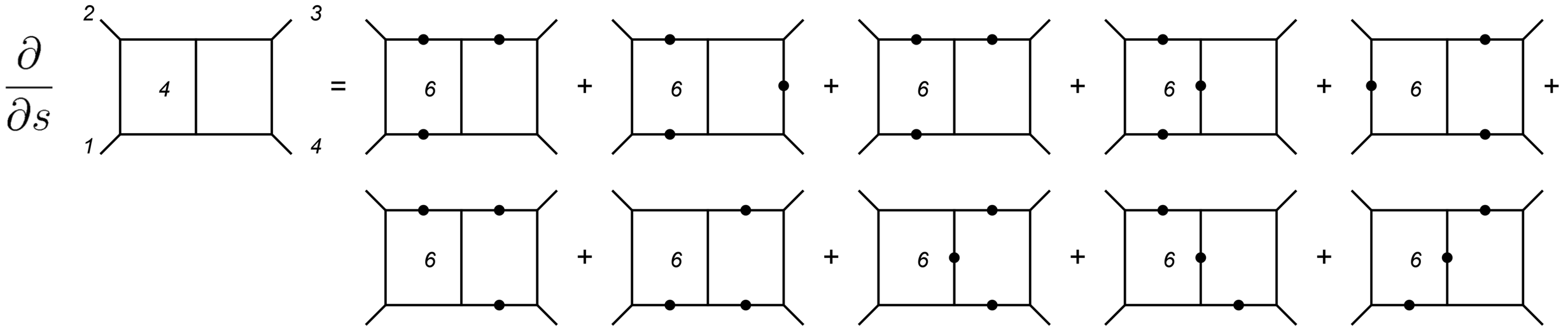}
 \end{center}\vspace{-0.2cm}
 \caption{Additional relation between two loop four point $d=4$ and $d=6$ double box integrals (their $\mathcal{I}$ functions). Note that only the last pair of $d=6$ integrals are individually dual conformal. }\label{dt_DBox}
 \end{figure}

Note that in this case the partial derivative with respect to $s$ will no longer give a simple answer but instead one obtains rather complicated identities between the $d=4$ double box integral and the sum of $d=6$ two loop integrals with $1/p^2$ and $1/(p^2)^2$ propagators, which is presented graphically in fig. \ref{dt_DBox}. The form of the $d=6$ part  is however totally defined by the form of $\mathcal{V}$ polynomial (\ref{AlphaPolyDBox}) of the $d=4$ part
The same relation is also valid in the Higgs regularization.

Remarkably, the $d=6$ integrals in such identities are not all manifestly dual conformally invariant. However, in the Higgs regularization dual conformal invariance is preserved for the  whole sum in a sense that both the RHS and the LHS of the relation are the functions of the  dual conformal cross ratios $u,v$ (\ref{DualConfRaitMassive}) and hence are annihilated by the action of $K^{\nu}$ in the form of (\ref{DualConfBoost_1}).

\subsection{Higher loops}

At three loops in $d=6-2\epsilon$ one has two possible topologies: the triple box and the tennis court ones. Correspondingly, one has more ways to distribute the $1/(p^2)^n$ propagators among the triple box and the tennis court rungs. Higher loops will bring even more possibilities. We do not systematically discuss them here but instead we point out that among different possible dual conformal integrals there always exist the one which is related to the $l$-loop $d=4$ ladder type diagram. Namely, one can see that for a diagram like this in $d=4$ the structure of the alpha representation polynomial $\mathcal{V}$ has the form (here, as before, we consider the horizontal ladder)
\begin{eqnarray}
\mathcal{V}=s P(\alpha_1,\ldots)+t \prod_{i=1}^{l+1}\alpha_i,
\end{eqnarray}
where $P$ is some (complicated) polynomial of $\alpha$ parameters and $\alpha_1,\ldots,\alpha_{l=1}$ are
the parameters associated with vertical rungs. Thus,  as in the previous cases we see that the following relation holds
\begin{equation}
D^{l}Box_4^{d=6}(s,t)=-\pi\left(t  \frac{\partial}{\partial t}-1\right)D^{l}Box_4^{d=4}(s,t).
\end{equation}
Here $D^{l}Box_4^{d=6}(s,t)$ is the $d=6$ $l$-loop dual conformal ladder type diagram with vertical rungs given by $1/(p^2)^2$ propagators and $D^{l}Box_4^{d=4}(s,t)$ is the $d=4$ $l$-loop dual conformal ladder. As for the derivative with respect to $\partial/\partial s$, one obtains similar identities between $d=4$ integral and the sum of $d=6$ integrals with $1/p^2$ and $(1/p^2)^2$ propagators as in the two loop example. The same will be true for the tennis court topology for any type of derivative. The structure of such identities, however, is  controlled by $\mathcal{V}$ polynomials of the corresponding $d=4$ diagrams.

In all such relations in the Higgs regularization dual conformal invariance is preserved in the same sense as in the two loop case. However, individual $d=6$ integrals may not be dual conformal invariant.

\section{Iterative structure of the integrals}

Now we are ready to discuss a possible iterative structure of the $d=6$ dual conformal integrals.
However first let us remind how the iterative structure manifests itself in the $d=4$ case.

In N=4 SYM for the colour ordered four point amplitude at the $l$-loop level the ratio $M_4^{d=4,(l)}=A_4^{d=4,(l)}/A_4^{d=4,(0)}$ is given by some combination of dual conformal integrals \cite{BDS1,BDS4loop,BDSVolovich}. For $l=1,2$ $M_4^{d=4,(l)}$ is given by
\begin{equation}
M_4^{d=4,(1)}=\frac{-1}{2}Box_4^{d=4}(s,t),~M_4^{d=4,(2)}=\frac{1}{4}\left(DBox_4^{d=4}(s,t)
+DBox_4^{d=4}(t,s)\right).
\end{equation}

It was pointed out in \cite{BDS1,BDS2loop1,BDS2loop2} that within dimensional regularization $M_4^{(1)}$ and $M_4^{(2)}$ are related with each other as:
\begin{equation}\label{BDS_at_2_loops}
M_4^{d=4,(2)}(\epsilon)=\frac{1}{2}\left(M_4^{d=4,(1)}(\epsilon)\right)^2+f^{(2)}(\epsilon)M_4^{d=4,(1)}(2\epsilon)
+C^{(2)}+O(\epsilon),
\end{equation}
where $f^{(2)}=a_1+\epsilon a_2+\epsilon^2a_3$, and $a_i,C^{(2)}$ are some known constants \cite{BDS1}.

Similar relations were found for higher loop orders for $M_4^{d=4,(l)}$, $l>2$ \cite{BDS1} which culminated in the BDS ansatz conjecture\footnote{$g=g_{YM}^2N_c/(16\pi^2)$ is the $d=4$ SYM perturbation theory expansion parameter.}
\begin{equation}\label{BDS_formula}
M_4^{d=4}=\sum_{l=0}^{\infty}g^lM_4^{d=4,(l)}=\exp \left[\sum_l g^l \bigg( f^{(l)}(\epsilon) M_4^{d=4,(1)}(l \epsilon)+C^{(l)}+{E^{(l)}_4(\epsilon)}\bigg)\right],
\end{equation}
where $f^{(l)}(\epsilon)$ has the same structure as before and is in fact related with cusp anomalous dimension  as $f^{(l)}(0)=\gamma_{l}/4$. $C^{(l)}$ is some constant and ${E^{(l)}_4(\epsilon)}\sim O(\epsilon)$.

Similar to the (\ref{1loopd4d6relation}) and (\ref{2loopd4d6relation}) this likely holds for the Higgs regularization as well. In \cite{Henn1,Henn2} it was shown that up to three loops the following relation holds (here for convenience we put all regulator masses equal $m_i=m$ so that the dual conformal cross ratios are equal to $u=m^4/s$ and $v=m^4/t$):
\begin{eqnarray}\label{BDS_formula_3}
\sum_{l=0}^{\infty}g^lM_4^{d=4,(l)}&=&\exp \left[ \frac{-\gamma_{cusp}(g)}{8}\left[L^2(u)+L^2(v)\right]-
\tilde{G}(g)\left[L(u)+L(v)\right]\right]\times\nonumber\\
&\times&\exp\left[\frac{\gamma_{cusp}(g)}{8}L^2\left(\frac{u}{v}\right)+\tilde{C}\right]+O(m^2).
\end{eqnarray}
Here the functions $\tilde{G}(g)$ and $\tilde{C}$ are different from their dimensional regularization counterparts but $\gamma_{cusp}(g)$ is the same \cite{Henn1}. We see that as expected the finite part of the amplitude $\exp(\gamma_{cusp}(g)L^2(u/v))=\exp(\gamma_{cusp}(g)L^2(s/t))$ is regularisation independent.
From this form one can obtain the evolution equation \cite{Korch_Conformal_properties,Henn2} of the following form:
\begin{eqnarray}\label{EvolEq3}
s\frac{\partial}{\partial s}log\left[\sum_{l=0}^{\infty}g^lM_4^{d=4,(l)}\right]=\omega(v),
%~\mbox{with}~
%\omega(v)=1+\frac{\gamma_{cusp}(g)}{4}L(v)-\tilde{G}(g).
%,~\mbox{with}~\omega(t)=1+\frac{\gamma_{cusp}(g)}{4}L(m^2/t)-\tilde{G}(g).
\end{eqnarray}
with:
\begin{eqnarray}
\omega(v)=1+\frac{\gamma_{cusp}(g)}{4}L(v)-\tilde{G}(g).
\end{eqnarray}
The function $\omega(v)$ is usually called the Regge trajectory.
%This equation will also hold in the Higgs regularization with exchange $\mu^2 \mapsto m^2$ \cite{Korch_Conformal_properties}.

\begin{figure}[ht]
 \begin{center}
 %\leavevmode
  \epsfxsize=11cm
 \epsffile{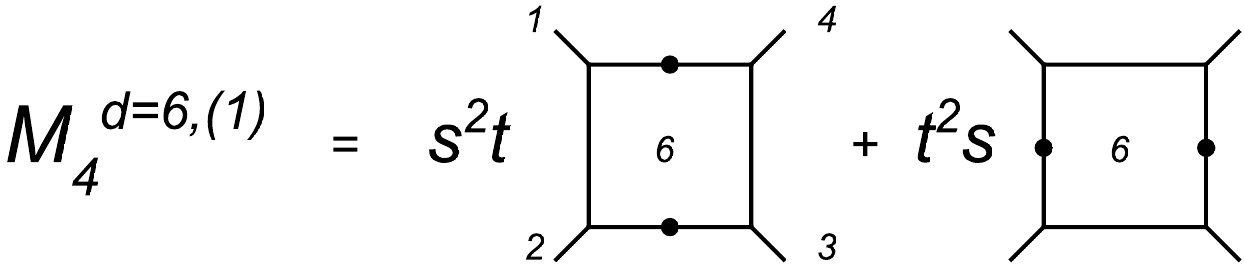}
 \end{center}\vspace{-0.2cm}
 \caption{The one loop amplitude candidate in $d=6$}\label{1loopAmpl}
 \end{figure}

Let us return to the $d=6$ case. Under the assumption that we are reconstructing the four point colour ordered amplitude in a hypothetical $d=6$ theory with dual conformal symmetry with massless particles, the only possible form for $M_4^{d=6,(l)}$ function at the one loop level is \cite{Lip} (see fig. \ref{1loopAmpl}):
\begin{equation}
M_4^{d=6,(1)}\sim Box_4^{d=6}(s,t)+Box_4^{d=6}(t,s).
\end{equation}

Using the results of the previous chapter, we conclude that the most general form (under the assumptions of section 2) for $M_4^{d=6,(2)}$ is:
\begin{equation}
M_4^{d=6,(2)}\sim\sum_{N=1}^{10}a_N \left(DBox_4^{d=6,N}(s,t)
+DBox_4^{d=6,N}(t,s)\right),
\end{equation}
where $a_N$ are some arbitrary numerical coefficients and $DBox_4^{d=6,N}(s,t)$ are the integrals of the form (\ref{eq::db61}) from table \ref{table1}. We then used our results from the previous section
and tried to fix the coefficients $a_N$ to satisfy relation (\ref{BDS_at_2_loops}). However, we found, that this is impossible for general values of $s$ and $t$. This observation can be interpreted in several ways:
\begin{itemize}
\item There is no iterative relation between $d=6$ four point dual conformal integrals.

\item The iterative relation between $d=6$ four point dual conformal integrals has a  more complicated form than (\ref{BDS_at_2_loops}).

\item The two loop basis of the dual conformal integrals is incomplete and one has to consider possible contribution from the integrals with the bubble and triangle subgraphs.
\end{itemize}

\begin{figure}[htb]
 \begin{center}
 %\leavevmode
  \epsfxsize=16cm
 \epsffile{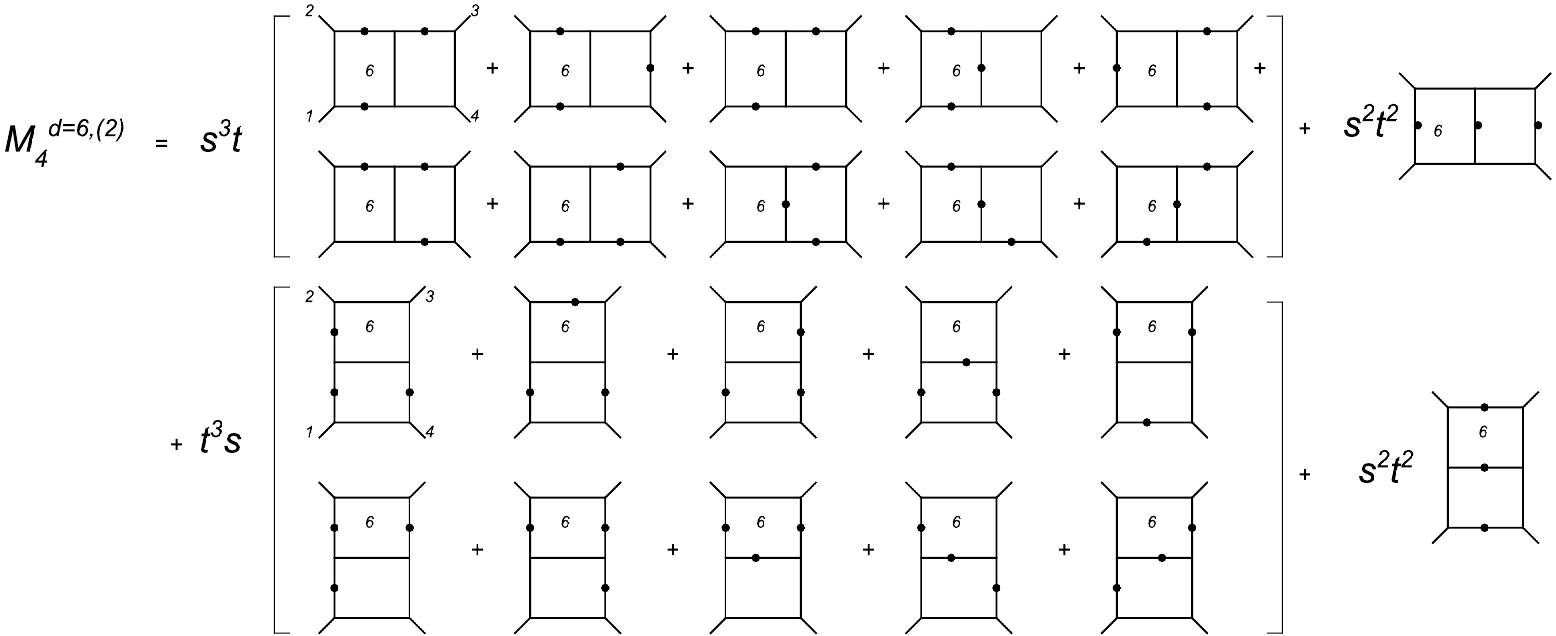}
 \end{center}\vspace{-0.2cm}
 \caption{The $d=6$ two loop amplitude candidates, which iterates the one loop result in fig.\ref{1loopAmpl}. The external momentum numeration is shown only on the first diagram in each channel.}\label{2loopAmpl}
 \end{figure}

Hereafter we will argue that some combination of the second and third propositions can be indeed realized.
To do so, we first consider the integrals in the Higgs regularization where the notion of the dual conformal invariance is more transparent. Taking into account $d=4$ vs $d=6$ relations between the integrals such as (\ref{1loopd4d6relation}) we
can combine together the sets  of $d=6$ integrals which we can be obtained from $d=4$ master integrals by the action of the operator
\begin{eqnarray}
\hat{\mathcal{D}}_{st}=s\frac{\partial}{\partial s}+t\frac{\partial}{\partial t}.
\end{eqnarray}
Let us label such sets of integrals as $\tilde M_4^{d=6,(l)}$. The sum of all $d=4$ integrals equals $M_4^{(l)}$ and equation (\ref{EvolEq3})
guarantees that the sum of the corresponding $d=6$ terms $\tilde M_4^{d=6,(l)}$ will be given by the $l$'th term of the expansion in $g$ of the BDS exponent multiplied by the prefactor given by the sum of Regge trajectories.  This gives us the relation
\begin{eqnarray}\label{D6theoryDefinition}
M_4^{d=6}&=&\sum_{l=0}^{\infty}{g_6}^l\tilde M_4^{d=6,(l)}=\sum_{l=0}^{\infty}{g_6}^l\left(\hat{\mathcal{D}}_{st}M_4^{d=4,(l)}-(l+1)M_4^{d=4,(l)}\right)\nonumber\\
&=&\left(-g\frac{\partial}{\partial g}-1+\omega(u)+\omega(v)\right)M_4^{d=4}\Bigg{|}_{g\mapsto {g_6}}+O(m^2).
\end{eqnarray}
Here $g_6$ is perturbation theory parameter in our hypothetical $d=6$ theory. The sets of the corresponding $d=6$ one and two loop integrals can be found in figs.
 \ref{1loopAmpl} and \ref{2loopAmpl} .

We want to stress once more that the resulting sum of $d=6$ integrals which one obtains from the action of the differential operator $\hat{\mathcal{D}}_{st}$  is uniquely determined by the form of $\mathcal{V}$ polynomial in the $\alpha$-representation. This effectively means that one can explicitly construct $\tilde M_4^{d=6,(l)}$ starting from the known expansion $M_4^{d=4,(l)}$ in terms of master integrals.

It is important to note that the operator $\hat{\mathcal{D}}_{st}$ does not break
dual conformal invariance and hence the RHS of relation (\ref{D6theoryDefinition}) is still the function of  dual conformal
cross ratios $u$ and $v$. However, individual $d=6$ integrals constructed from $1/p^2$ and $1/(p^2)^2$ propagators may not be dual conformal invariant.
Therefore, if we want to construct the combination of $d=6$ integrals with iterative structure we likely have to abandon the  requirement that individual $d=6$ integrals obey dual conformal invariance. It  is restored in total combination of such integrals. This is in fact rather well-known phenomenon when individual Feynman diagrams (integrals) contributing to some amplitude may have less symmetry than the amplitude itself.

An interesting question to ask is whether it is possible, for example at the two loop level, to express the combination of the integrals which are individually not dual conformal invariant via the basis of dual conformal integrals in $d=6$ ? We will avoid a detailed discussion here, and only mention that it is likely that the basis of (\ref{table1}) integrals regularized with Higgs regularization will not be sufficient and integrals different from the double box topology will be required. See fig. \ref{HiggsReg} B) as an example.

It seems that  similar iterative structure appears also in the case of dimensional regularization with
exchange of the Regge trajectory $\omega(v)$ computed in the Higgs regularization with $\omega(t)$:
\begin{eqnarray}\label{ReggeTrInDimReg}
\omega(t)=\frac{\gamma_{cusp}(g)}{4}L\left(\frac{\mu^2}{t}\right)+\frac{G(g)}{2}+\frac{1}{2\epsilon}\int_0^g\frac{dg'}{g'}\gamma_{cusp}(g')+O(\epsilon).
\end{eqnarray}
Note, however, that for practical implementation of this formula one
 needs to know the $O(\epsilon)$ terms in the definition of $\omega$ due to the interference with the
poles in $\epsilon$ from the BDS exponent. This is why we made emphasis on the Higgs regularization where such problem is absent (analogous terms $m~log(m)$ in the Higgs regularization can be simply omitted).
%Another important question is whether the analogue of (\ref{DualConformalAnomaly1}) will hold for $M_4^{d=6}$ in dimensional regularization. At the first glance the answer is positive but more detailed investigation is needed to make some definitive statement.

Note also that in this form the iterative structure is consistent with the $AdS$ volume calculation of \cite{Lip} in a sense that the sum of the $M_4^{d=6,(l)}$ terms over $l$ will be proportional to
\begin{eqnarray}\label{ReggeTrInDimReg}
\sum_{l=0}^{\infty}M_4^{d=6,(l)}\sim exp\left(\mathcal{S}_{s}+\mathcal{S}_{t}
+\frac{\gamma(g_4)}{8}L^2\left(\frac{s}{t}\right)\right)\Bigg{|}_{g\mapsto g_6},
\end{eqnarray}
However, in our construction the proportionality coefficient is a non-trivial pre-exponent function containing the $1/\epsilon$ pole and terms proportional to $\epsilon$. This situation is reminiscent of the Schwinger pair production process where the exponential factor can be predicted from quasi-classical considerations, while the pre-exponent can be obtained only from full computations. It is also interesting to compare this result with investigation of  \cite{ABJM1,ABJM2} where similar connection between $d=4$ N=4 SYM and $d=3$ ABJM amplitudes where discovered.

\section{Conclusion}
In this article we investigated the conjecture of \cite{Lip} on a possible iterative structure of $d=6$
four point (pseudo) dual conformal integrals. By construction such integrals contain $1/p^2$ as well as $1/(p^2)^2$ propagators. We found by explicit computations of the set of two loop integrals presented in \cite{Lip} that there is no iterative pattern between such integrals similar to their $d=4$ counterparts (the BDS exponent). Alternatively here we propose the  algorithm how to construct such a set of $d=6$ integrals with $1/p^2$ and $(1/p^2)^2$ propagators which possess an iterative structure that is consistent with the AdS minimal volume computations of \cite{Lip}.  Individual integrals in these sets, however, are not explicitly dual conformal invariant and the  invariance in appropriate regularization is restored  only for the whole set of integrals.

\section*{Acknowledgements}
The authors are grateful to A.I.Onishchenko for useful discussions.
This work was supported by the Russian Science Foundation (RSF) grant \#16-12-10306.

\section*{Appendix}

\section*{1-loop Box diagrams}
In the off-shell case the $l$-rung $d=4$ ladder diagram is given by the
Davydychev-Ussyukina function $\Phi_4^{(l)}(u,v)$:
\begin{eqnarray}
%\label{generalUDf}
D^lBox(s,t)=\Phi_4^{(l)}(u,v),
\end{eqnarray}
with
\begin{eqnarray}
\label{generalUDf}
\Phi_4^{(L)}(u,v)&=&\frac{1}{\lambda}\bigg(\frac{1}{L!}\sum_{j=L}^{2L}\frac{j!L(v/u)^{2L-j}}{(j-L)!(2L-j)!} \left(\text{Li}_{j}(-\rho u)+(-1)^j\text{Li}_{j}(-\rho v)\right)+\nonumber\\
 &+&2\sum_{k=0}^{'L}\sum_{l=0}^{L}
 \frac{(k+l)!(1-2^{1-k-l})}{k!l!(L-k)!(L-l)!}L(\rho v)^{L-k}L(\rho u)^{L-l}\zeta(k+l) \bigg),
\end{eqnarray}
where $\sum^{'}$ corresponds to selection condition on $l$ and $k$: $k+l=\mbox{even}$,
$\rho=2/(1-u-v-\lambda)$, $\lambda=(1-u-v)^2-4uv)^{1/2}$, and $u,v$ are given by (\ref{Cross1}).
In the $l=1$ case in the small $m^2$ limit (here $m^2=p^2_i$) this expression can be simplified to:
\begin{eqnarray}
\label{UDf_1loop}
Box_4^{d=4}(s,t)&=&-\frac{1}{2}L\left(\frac{m^4}{st}\right)^2+O(m^2).
\end{eqnarray}

In Higgs regularisation $d=4$ box diagram is given by:
\begin{eqnarray}
\label{}
Box_4^{d=4}(s,t,m)&=&2L(u)L(v)-\pi^2+O(m^2),
\end{eqnarray}
with $u,v$ are given by (\ref{DualConfRaitMassive}). Note that here $m^2$ is the mass parameter in the propagators of box integral in contrast to (\ref{UDf_1loop}).

For $d=6$ as we mentioned above one-loop four-point diagram (\ref{d6Box}) in dimensional regularisation can be calculate via MB-representation. The result is given by (hereafter we as usual add $\exp(+\gamma_E\epsilon)$) factor to each loop integral:
%\begin{equation}
% B_{k,1}^6(x,\epsilon)= \sum_n \frac{c_n}{\epsilon^n}
% \end{equation}
 \begin{equation}
 Box_4^{d=6}(s,t)=\left(\frac{\mu^2}{s}\right)^{\epsilon}\sum_{n=2}^{-\infty}\frac{c_n(x)}{\epsilon^n},
 \end{equation}
 with ($x=-s/t$)
\begin{multline}
 c_{2}=4; c_{1}=-2 (\ln (x)-1);  c_{0}=-8 \zeta (2); c_{-1}=\frac{1}{6} \left(-42 \zeta (2)-68 \zeta (3)+42 \zeta (2) \ln (x)+\right. \\ \left.+2 \ln (x)^3-6 \ln (x)^2\right)+ \left(-6 \zeta (2)-2 \text{Li}_3(-x)+2 \text{Li}_2(-x) \ln (x)+6 \zeta (2) \ln (x+1)-\right. \\ \left.-\ln (x)^2+\ln (x)^2 \ln (x+1)\right)+\left(\frac{  \ln (x)^2+6 \zeta (2) }{x+1}+2 x   (\ln (x)-1)\right).
 \end{multline}

For completeness let us also write $d=4$ dimensionally regularised Box integral, which is equal to:
  \begin{equation}
  Box_4^{d=4}(s,t)=\left(\frac{\mu^2}{s}\right)^{\epsilon}\sum_{n=2}^{-\infty}\frac{c_n(x)}{\epsilon^n},
  \end{equation}
  with ($x=-s/t$)
 \begin{multline}
  c_{2}=4; c_{1}=-2\ln (x);  c_{0}=-8 \zeta (2); c_{-1}=\frac{1}{6} \left(-68 \zeta (3)+42 \zeta (2) \ln (x)+2 \ln (x)^3\right) \\ + \left(-6 \zeta (2)-2 \text{Li}_3(-x)+2 \text{Li}_2(-x) \ln (x)+6 \zeta (2) \ln (x+1)+\ln (x)^2 \ln (x+1)\right).
  \end{multline}

\section*{2-loop box diagrams}
For the double boxes we present only the pole parts. As in the previous section we present our answers in the form
\begin{equation}
DBox_4^{d=6,N}(s,t)=\left(\frac{\mu^2}{s}\right)^{2\epsilon}\sum_{n=2}^{-\infty}\frac{c_n(x)}{\epsilon^n},
\end{equation}
with $x=-s/t$.
For $N=1$ (see table \ref{table1}) we have
\begin{multline}
c_{4}=4; \;
c_{3}=-5\left(\ln(x)-1\right); \;
c_{2}=-\frac{1}{2}\left(-30\zeta(2)+4\ln^2(x)-8\ln(x)\right);\\
c_{1}=\frac{1}{6 }\left(198\zeta(2)\ln(x)-126\zeta(2) +4\ln^3(x)-72\zeta(2)\ln(1+x) -12\ln^2(x)\ln(1+x)-\right. \\
            \left.-24\ln(x)\mathrm{Li}_2(-x)+48\mathrm{Li}_2(-x)+24\mathrm{Li}_3(-x)-130 \zeta(3)\right) -\frac{2\left(6\zeta(2)+\ln^2(x)\right)}{(1+x)} \bigg).
\end{multline}

 For $N=2$ we obtain:
\begin{multline}
c_{4}=4; \; c_{3}=7-5 \ln (x); \; c_{2}=-144 \zeta (2)+18 \ln (x)^2-90 \ln (x)+35 ~; \\ c_{1}=\frac{1}{6 (x+1)}\left(-138 \zeta (2)-130 \zeta (3)-24 x \text{Li}_3(-x)-24 \text{Li}_3(-x)+24 x \text{Li}_2(-x) \ln (x)+\right. \\ \left.+24 \text{Li}_2(-x) \ln (x)-66 \zeta (2) x-130 \zeta (3) x+198 \zeta (2) x \ln (x)+198 \zeta (2) \ln (x)+\right. \\ \left.+72 \zeta (2) x \ln (x+1).+72 \zeta (2) \ln (x+1)-6 x+4 x \ln (x)^3+4 \ln (x)^3+48 x \ln (x)^2+\right. \\ \left.+12 x \ln (x+1) \ln (x)^2+12 \ln (x+1) \ln (x)^2+36 \ln (x)^2-60 x \ln (x)-60 \ln (x)-6\right).
\end{multline}

For $N=3$ we get (here we present also the finite part)
\begin{multline}
c_{4}=0; c_{3}=\frac{1}{2}; \; c_{2}=\frac{\ln (x)}{2}; \; c_{1}=\frac{6 \zeta (2)+\ln (x)+1}{2 }; \;
c_{0}=-\frac{1}{96} \left(-384 \zeta (2)+1184 \zeta (3)-\right. \\ \left.-1152 \zeta (2) x -624 \zeta (2) \ln (x)-192 x-32 \ln (x)^3-\right. \\ \left.-192 x \ln (x)^2-96 \ln (x)^2+ 192 x \ln (x)-144 \ln (x)-27\right).
\end{multline}

Integrals $N=4$ and $N=5$ are the same end evaluates to
\begin{equation}
c_{4}=0; c_{3}=-2; \; c_{2}=4 \ln (x)-3; \; c_{1}=\frac{78 \zeta (2)+12 \ln (x)}{3}.
\end{equation}

For $N=6$ and $N=7$ integrals the answers are the same:
\begin{equation}
c_{4}=0; c_{3}=\frac{1}{2}; \; c_{2}=\frac{-8 \ln (x)-7}{16}; \; c_{1}=\frac{-38 \zeta (2)+12 \ln (x)-27}{16}.
\end{equation}

Integrals $N=8$ and $N=9$ also evaluates to the identical answers:
\begin{equation}
c_{4}=0; c_{3}=4; \; c_{2}=-5 (\log (x)-1); \; c_{1}=\frac{1}{6}(-24 x+12 \log ^2(x)-24 \log (x)-102 \zeta(2) ^2-21).
\end{equation}

And finally for $N=10$ we got
\begin{equation}
c_{4}=0; c_{3}=1; c_{2}=-\ln (x);  c_{1}=\frac{-26 \zeta (2)- \ln (x)-17}{4}.
\end{equation}

\bibliographystyle{hieeetr}
\bibliography{refs_d6}

\end{document}